\documentclass[fleqn,10pt]{wlscirep}
\usepackage[utf8]{inputenc}
\usepackage[T1]{fontenc}
\usepackage[flushleft]{threeparttable}
\usepackage[sort, numbers]{natbib}
\usepackage{xcolor}
\usepackage{lineno}
\usepackage{multirow}
\usepackage{tikz,xcolor,hyperref}

\definecolor{lime}{HTML}{A6CE39}
\DeclareRobustCommand{\orcidicon}{%
	\begin{tikzpicture}
	\draw[lime, fill=lime] (0,0) 
	circle [radius=0.16] 
	node[white] {{\fontfamily{qag}\selectfont \tiny ID}};
	\draw[white, fill=white] (-0.0625,0.095) 
	circle [radius=0.007];
	\end{tikzpicture}
	\hspace{-2mm}
}

\foreach \x in {A, ..., Z}{%
	\expandafter\xdef\csname orcid\x\endcsname{\noexpand\href{https://orcid.org/\csname orcidauthor\x\endcsname}{\noexpand\orcidicon}}
}

\usepackage{censor}
\StopCensoring 

\title{Avoiding methane emission rate underestimates when using the divergence method}

\author[1,*]{\orcidA{}\blackout{Clayton Roberts}} 
\author[2]{\blackout{Rutger IJzermans}}
\author[2]{\orcidF{}\blackout{David Randell}}
\author[2]{\blackout{Matthew Jones}}
\author[3,4]{\orcidD{}\blackout{Philip Jonathan}}
\author[1,5,6]{\orcidC{}\blackout{Kaisey Mandel}}
\author[7]{\orcidE{}\blackout{Bill Hirst}}
\author[1,8]{\orcidB{}\blackout{Oliver Shorttle}}

\affil[1]{\blackout{Institute of Astronomy, University of Cambridge, Madingley Road, Cambridge CB3 0HA, United Kingdom}}
\affil[2]{\blackout{Shell Global Solutions International B.V., Grasweg 31, 1031 HW Amsterdam, The Netherlands}}
\affil[3]{\blackout{Department of Mathematics and Statistics, Lancaster University, Lancaster LA1 4YW, United Kingdom}}
\affil[4]{\blackout{Shell Research Ltd, London SE1 7NA, United Kingdom}}
\affil[5]{\blackout{Statistical Laboratory, Department of Pure Mathematics and Mathematical Statistics, University of Cambridge, Wilberforce Road, Cambridge CB3 0WB, United Kingdom}}
\affil[6]{\blackout{The Alan Turing Institute, Euston Road, London NW1 2DB, United Kingdom}}
\affil[7]{\blackout{Atmospheric Monitoring Sciences, Amsterdam, The Netherlands}}
\affil[8]{\blackout{Department of Earth Sciences, University of Cambridge, Downing Street, Cambridge CB2 3EQ, United Kingdom}}

\affil[*]{\blackout{cnr31@cam.ac.uk}}


\begin{abstract}
Methane is a powerful greenhouse gas, and a primary target for mitigating climate change in the short-term future due to its relatively short atmospheric lifetime and greater ability to trap heat in Earth's atmosphere compared to carbon dioxide. Top-down observations of atmospheric methane are possible via drone and aircraft surveys as well as satellites such as the TROPOspheric Monitoring Instrument (TROPOMI). Recent work has begun to apply the divergence method to produce regional methane emission rate estimates. Here we show that when the divergence method is applied to spatially incomplete observations of methane, it can result in negatively biased time-averaged regional emission rates. We show that this effect can be counteracted by adopting a procedure in which daily advective fluxes of methane are time-averaged before the divergence method is applied. Using such a procedure with TROPOMI methane observations, we calculate yearly Permian emission rates of 3.1, 2.4 and 2.7 million tonnes per year for the years 2019 through 2021. We also show that highly-resolved plumes of methane can have negatively biased estimated emission rates by the divergence method due to the presence of turbulent diffusion in the plume, but this is unlikely to affect regional methane emission budgets constructed from TROPOMI observations of methane. The results from this work are expected to provide useful guidance for future implementations of the divergence method for emission rate estimation from satellite data -- be it for methane or other gaseous species in the atmosphere. 
\end{abstract}

\begin{document}

\flushbottom
\maketitle
%
%
\thispagestyle{empty}

\section{Introduction}
Methane is a powerful greenhouse gas, with a far greater warming potential (84 times greater on a 20-year timescale) and shorter atmospheric lifetime (9 years instead of centuries) than carbon dioxide \cite{IPCC_2021, Balcombe_2018}. These attributes make methane an attractive target for mitigating the short-term effects of climate change, and have been the focus of recent climate summits and global commitments towards emission reductions \cite{Global_Methane_Pledge}. Nevertheless, in recent years, the rate of increase of atmospheric methane has itself increased. \cite{NOAA_CH4, Peng_2022}. Roughly 30\% of anthropogenic methane emissions are attributed to the fossil fuel industry \cite{IEA_2022, Saunois_2020}, making increased monitoring and accounting of emissions from this sector an important factor in meeting national commitments towards methane emission reductions. \\

Satellite observations are a powerful tool for monitoring atmospheric methane abundances \cite{Jacob_2016}, with remote sensing of methane from space providing opportunities for repeated and unscheduled monitoring of emissions. The era of greenhouse gas observing satellites began with the SCanning Imaging Absorption SpectroMeter for Atmospheric CHartographY (SCIAMACHY) \cite{Bovensmann_1999} in 2003, and subsequent generations of satellites have given rise to instruments with ever increasing capabilities. The TROPOSpheric Monitoring Instrument (TROPOMI) provides daily global coverage of methane observations with an updated 5.5x7 km$^2$ pixel resolution \cite{Veefkind_2012}, whilst other instruments such as GHGSat provide intermittent, targeted methane observations down to 50x50 m$^2$ resolution \cite{Jervis_2021}. Many greenhouse gas-observing satellites lack the spatial resolution to calculate asset-level emissions, and so aircraft and drone surveys are used to bridge this gap. These instruments can image and estimate facility-level methane emission rates \cite{Cusworth_2021, Ayasse_2022}, and such facility-level measurements can be used for reporting under the Oil \& Gas Methane Partnership 2.0 (OGMP 2.0) framework. This is a multi-stakeholder initiative launched by the United Nations Environment Programme (UNEP) and the Climate and Clean Air Coalition, aimed at improving the accuracy and transparency of methane emissions reporting in the oil and gas sector \cite{OGMP}. Recently, ``top-down" methane emission estimates calculated from aircraft observations have been found to be in disagreement to ``bottom-up" emission estimates reported from industrial activity \cite{Zavala_2015, Alvarez_2018}, and more work is required to reconcile these standards of reporting methane emissions. \\

There are a variety of methods for constructing top-down emission estimates from satellite observations of methane. Some analyses use Bayesian methods to estimate regional methane emission rates \cite{Zhang_2020}; Bayesian methods require priors to be specified on the spatial distributions of emission rates, which are sometimes constructed from bottom-up emission estimates. Other methods such as the inverse Gaussian plume method \cite{Bovensmann_2010, schwandner_spaceborne_2017, nassar_quantifying_2017}, the source pixel method \cite{Jacob_2016,buchwitz_satellite-derived_2017}, the cross-sectional flux method \cite{Schneising_2020,white_formation_1976}, and the integrated mass enhancement (IME) method \cite{frankenberg_airborne_2016,thompson_space-based_2016} rely directly on column measurements of methane (either remotely sensed or in-situ) and wind field data. These methods are particularly attractive for construction top-down emission budgets as they are entirely divorced from prior estimates that are informed by self-reported production statistics.\\

Another top-down methodology for emission estimation is the divergence method \cite{Beirle_2019}. Like the inverse Guassian plume, source pixel, cross-sectional flux and IME methodologies, the divergence method relies solely on wind field and column observations of the quantity of interest. In contrast to these other top-down methodologies, the divergence method can also be used for regional-scale emission rate estimation \cite{Beirle_2019, Liu_2021, Veefkind_2023}. In the divergence method, emission is calculated via the spatial gradients of zonal (i.e., advective) fluxes. Mathematically, the total sources and sinks of emission $E$ are calculated via the emission equation

\begin{equation}\label{eq:simple_emission_eq}
    E = \nabla\cdot\vec{F}^\mathrm{adv} \quad [\mathrm{kg}\,\mathrm{m}^{-2}\mathrm{s}^{-1}],
\end{equation}

\noindent where $\vec{F}^\mathrm{adv}$ is the advective flux of a quantity of interest (e.g., methane or other pollutants). The divergence method for estimating the spatial distribution of methane emissions is attractive because it is entirely data-driven and does not rely on prior estimates of spatial emission distributions as extensively as Bayesian methods do. Unlike the source pixel method, the divergence utilises wind field and concentration information away from source locations \cite{Varon_2018}. Originally presented as a method for estimating the location and emission rates of sources of nitrogen dioxide, this methodology is now being used to estimate regional-level methane emission rates \cite{Liu_2021, Veefkind_2023}. \\

It is important to make explicit some important assumptions that are currently intrinsic to this methodology. First, plumes of any gas (including methane) propagate through the atmosphere not only by advection, but also molecular and turbulent diffusion \cite{Stockie_2011,Taylor_1922,Roberts_2002}; in atmospheric transport, turbulent diffusion is usually the dominant effect over molecular diffusion in practice. The emission equation $E = \nabla\cdot\vec{F}^\mathrm{adv}$ that is central to many regional-level methane budget estimates does not take the effect of turbulent diffusion into account. Correcting for turbulent diffusion requires the usage of a modified emission equation

\begin{equation}\label{eq:modified_emission_eq}
    E = \nabla\cdot\left(\vec{F}^\mathrm{adv} - \vec{F}^\mathrm{dif}\right)  \quad [\mathrm{kg}\,\mathrm{m}^{-2}\mathrm{s}^{-1}],
\end{equation}

\noindent where $\vec{F}^\mathrm{dif}$ is the turbulent diffusive flux of a quantity of interest. Second, regional methane emission rate estimates are often time-averaged. The linear property of the divergence operator means that time-averaged estimated emission rates could be calculated in two different ways: first, by time-averaging daily estimated emission rates, or second, by taking the divergence of time-averaged daily advective fluxes. With perfect observations, these procedures should yield identical results, but satellite observations of methane are often spatially disrupted with missing coverage. To the best of our knowledge, no work has yet been done to examine the consequence of this choice of order of operations when the divergence method is applied to methane observations. \\

 In this work, we derive analytical expressions for $\vec{F}^{\mathrm{dif}}$, and generate synthetic simulated satellite observations of Gaussian plumes to examine under what physical scenarios it becomes important to include $\vec{F}^{\mathrm{dif}}$ in the emission equation. We find that when a plume of methane is relatively diffusive, methane emission estimates via the divergence method can become inaccurate if $\vec{F}^{\mathrm{dif}}$ is excluded from the divergence calculation (under certain conditions). In this case, the estimated emission rate of the source is underestimated, and the spatial distribution of emissions is incorrectly distributed. We also demonstrate that time-averaged emission estimates calculated from spatially incomplete observations may be inaccurate if care is not taken to use a time-averaged advective flux in the emission equation (as opposed to taking the time-average of daily emission estimates via the divergence method). We compare the results of our synthetic study to a case study of the Permian basin, using TROPOMI observations of methane \cite{TROPOMI_CH4_DATA}. We find that it is unlikely that a regional methane emission budget calculation of the Permian via the divergence method would be negatively biased due to the effects of any turbulent diffusion. We do find though that the sparse nature of the TROPOMI methane data product can result in negatively biased time-averaged methane emission rate estimates when the divergence method is used, specifically affecting cases where  daily emission rate estimates which are time-averaged to produce the time-averaged emission rate estimate. When using the divergence method in conjunction with spatially sparse observations, it is important to take the divergence of time-averaged daily fluxes to obtain a time-averaged emission estimate. 

\section{Results}
\subsection{Synthetic case study}
\begin{figure}[ht]
\centering
\includegraphics[width=7.2in]{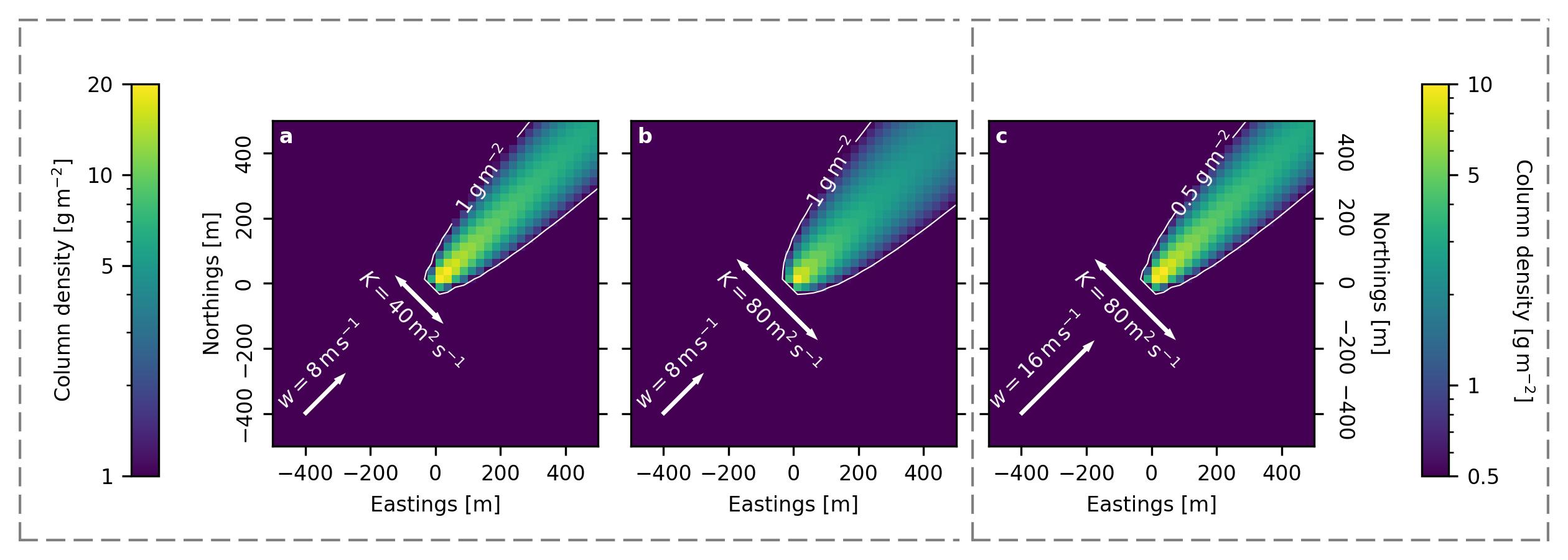}
\caption{Simulated column-integrated ideal Gaussian plumes. All the plumes shown have the same emission rate of $Q_\mathrm{true} = 10\,\mathrm{kg}\,\mathrm{s}^{-1}$ and wind angle $\theta=45$ degrees. \textbf{a} Simulated plume observation with $w=8\,\mathrm{m}\,\mathrm{s}^{-1}$ and $K=40\,\mathrm{m}^2\,\mathrm{s}^{-1}$. \textbf{b} Same simulated observation as in \textbf{a}, this time with doubled $K$. \textbf{c} As in \textbf{a}, this time with doubled $K$ and doubled $w$. Note that panels \textbf{a} and \textbf{b} are plotted with the same colorscale with the colorbar shown on the left hand side, and that panel \textbf{c} is plotted with the colorbar on the right hand side. The simulated plume in \textbf{b} has a wider opening angle than in \textbf{a} due to the increased ratio of $K$ to $w$. The plume in \textbf{c} has the same opening angle as the plume in \textbf{a} because $K/w=5\,\mathrm{m}$ for both plumes. However, the column density of the plume in \textbf{c} is half that of the column density in \textbf{a}, because according to Eq. \ref{eq:plume_generation_equation}, column density scales as $\sim \left(K\,w\right)^{-1/2}$. } 
\label{fig:1}
\end{figure}

\noindent We generate simulated satellite observations of ideal steady-state Gaussian plumes \cite{Stockie_2011} resulting from isolated point sources with known emission rates, and use them in an investigative synthetic case study. These plumes are characterised by a known emission rate $Q_{\mathrm{true}}$ [kg s$^{-1}$], wind speed $w$ [m s$^{-1}$], wind angle $\theta$ relative to the x-axis of the observation grid, and constant of turbulent diffusion $K$ [m$^2$ s$^{-1}$]. Simulated plume observations are generated via Eq. \ref{eq:plume_generation_equation}, which is derived in S1. Note that when we calculate grid cell values, we numerically spatially integrate Eq. \ref{eq:plume_generation_equation} over the area of the grid cell, and so the results of our synthetic study are independent of grid cell resolution. This is important to bear in mind as real observational instruments will have pixel resolutions spanning from meter to kilometer scales, which changes the amount of fine-scale structure that will be visible in observed plumes. Fig. \ref{fig:1} shows some examples of how these parameters affect plume morphology. For our simulated observations, we use the divergence method to estimate the spatial emission field both with and without including the diffusive flux term in the emission equation. We find in our synthetic study that there are a variety of scenarios where the use of the divergence method results in negatively biased estimated emission rates, i.e., when turbulent diffusion is neglected in some cases, or when time-averaged estimated emission rates are calculated without time-averaging daily flux fields when daily observations are spatially incomplete. \\

\subsubsection{Underestimating emission rates due to turbulent diffusion in the plume}\label{sec:synthetic_study_turbulent_diffusion}
\begin{figure}[ht]
\centering
\includegraphics[width=7.2in]{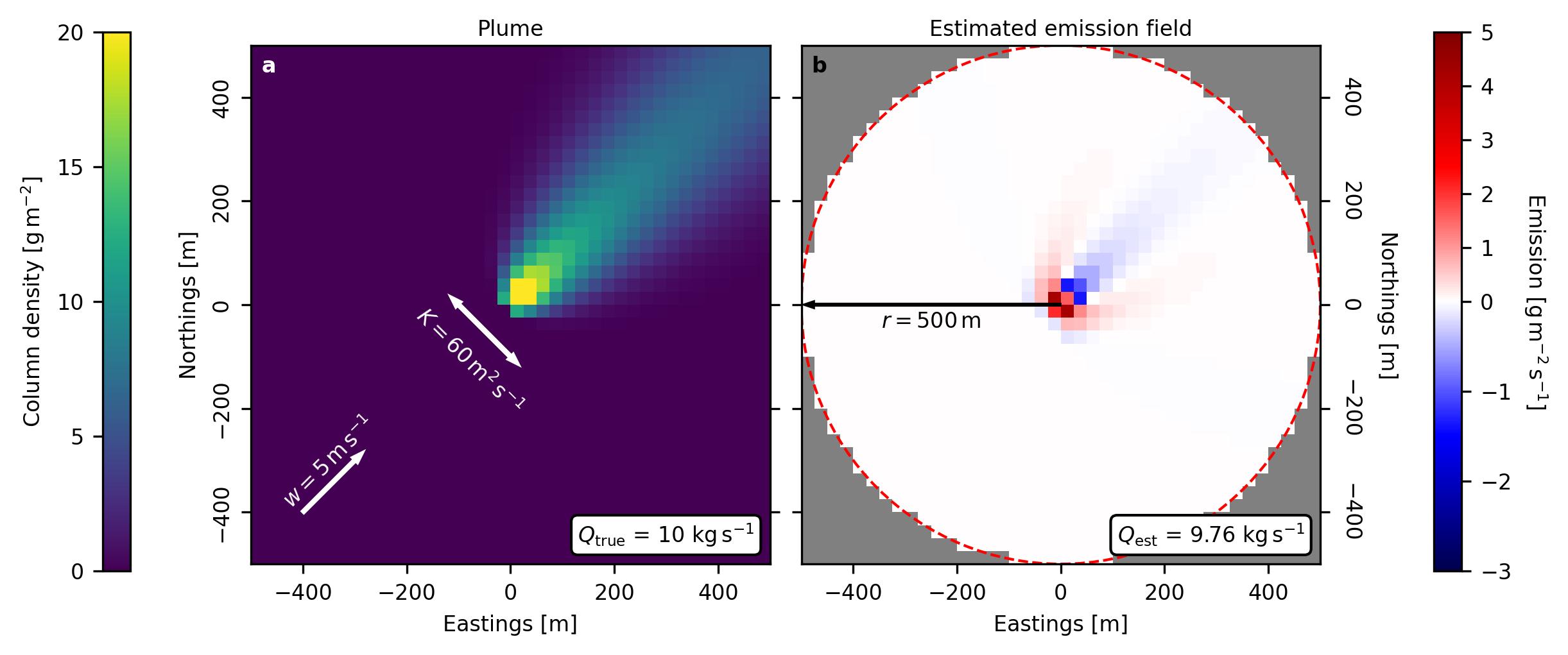}
\caption{\textbf{a} A simulated column-integrated Gaussian plume, with wind angle $\theta=45$ degrees and other parameters as indicated on the plot. \textbf{b} The estimated emission field of the plume using the divergence method and emission equation $E = \nabla\cdot\vec{F}^\mathrm{adv}$. As the plume in \textbf{a} was constructed using a point source of positive emission located at the origin, we can see in \textbf{b} that the estimated emission field has an incorrect spatial distribution. ``Positive'' emissions are incorrectly distributed in a horseshoe-shaped pattern around the origin, and ``negative'' emissions (or sinks) downwind in the shadow of the plume. When the estimated emission field is integrated over a circular region of radius $r=500$ m centered on the origin, the total emission rate is slightly underestimated. For a fixed $r$ and true point source emission rate $Q_\mathrm{true}$, the total estimated emission rate $Q_\mathrm{est}$ will be underestimated to a greater extent if $K/w$ is increased. When $K/w$ increases for the plume, the plume becomes more ``diffuse'', and the emission equation $E = \nabla\cdot\vec{F}^\mathrm{adv}$ becomes less able to explain the resulting column density as arising from a point source of emission.}
\label{fig:2}
\end{figure}

Fig. \ref{fig:2} demonstrates the application of the divergence method to a simulated satellite observation of a plume. We begin by using the standardly formulated emission equation where the turbulent diffusive flux term term is not included, i.e., 

\begin{align}
    E &= \nabla\cdot\vec{F}^\mathrm{adv} \quad [\mathrm{kg}\,\mathrm{m}^{-2}\mathrm{s}^{-1}]\label{eq:stand} \\
    \vec{F}^\mathrm{adv} &= C\,\vec{w} = C\,w_x\,\vec{e_x} + C\,w_y\,\vec{e_y} \quad [\mathrm{kg}\,\mathrm{m}^{-1}\,\mathrm{s}^{-1}], 
\end{align}

\noindent where $C$ is the simulated column density of methane in a grid cell and $\vec{w}$ is the simulated wind vector in a grid cell. We find that when estimating the emission field without including the turbulent diffusion term in the emission equation, emissions are incorrectly spatially distributed in the estimated emission field. Rather than estimating a single point of positive emission, we instead find that ``arrowhead" shapes of positive emission are estimated, in conjunction with negative emissions (or sinks) downwind within the plume shadow (see Fig. \ref{fig:2}\textbf{b}). To obtain a total estimated emission rate $Q_{\mathrm{est}}$, we spatially integrate the estimated emission field within a circle of fixed radius centered on the source location. Using Eq. \ref{eq:stand}, we find that the total estimated emission rate of the emission field is underestimated, the extent of which depends on the conditions discussed in the following two paragraphs.  \\

When increasing the opening angle of a plume (which scales as a function of $K/w$), the total estimated emission rate decreases for a fixed radius of spatial integration over the estimated emission field. This demonstrates that the emission rates of ``diffuse'' plumes are poorly estimated when using the divergence method, relying solely on advective fluxes. To first order, the underestimation is proportional to $K/w$. It is important to note in our synthetic study that we measure estimated emission rates as percentages of the true emission rate, and so results are independent of the mass emission rate of the point source. For a real instrument, higher emission rates mean a higher measured signal-to-noise ratio.\\

\begin{figure}[ht]
\centering
\includegraphics[width=7.2in]{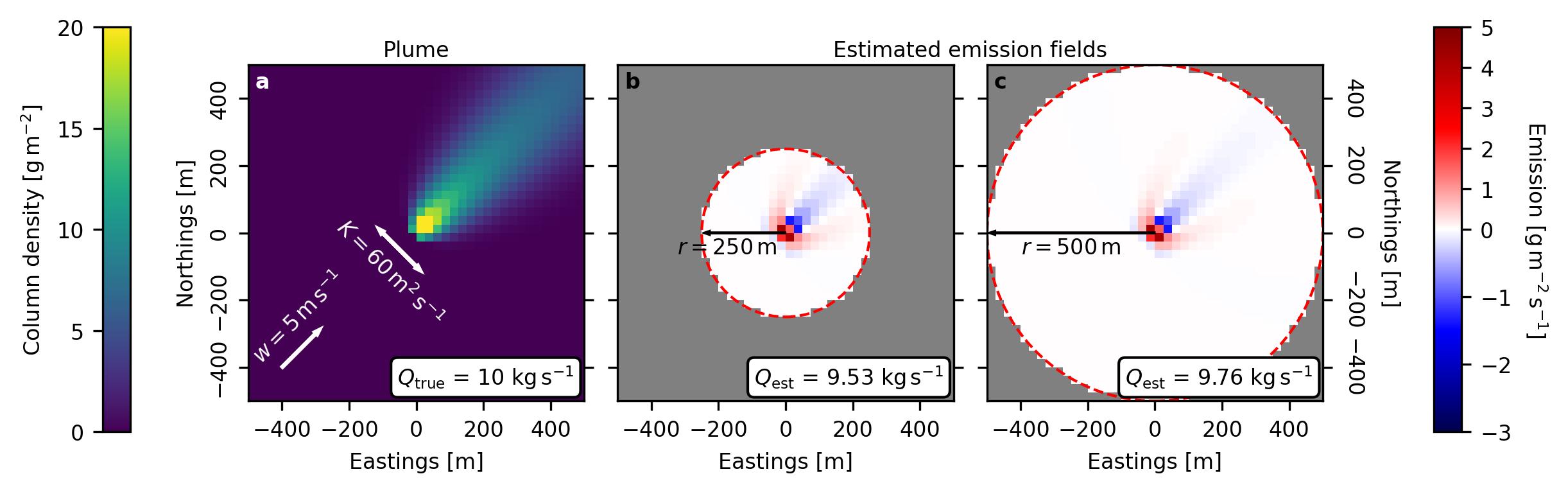}
\caption{\textbf{a} A simulated column-integrated Gaussian plume. \textbf{b} The estimated emission field using the divergence method, integrating over a circular area of radius $r=250$ m to obtain the total estimated emission rate $Q_{\mathrm{est}}$. \textbf{c} The same estimated emission field as in \textbf{b}, this time integrated over a circular region with $r=500$ m to obtain $Q_{\mathrm{est}}$. As $r$ increases, $Q_\mathrm{est}$ is underestimated to a lesser extent.}
\label{fig:3}
\end{figure}

\begin{figure}[ht]
\centering
\includegraphics[width=3.5in]{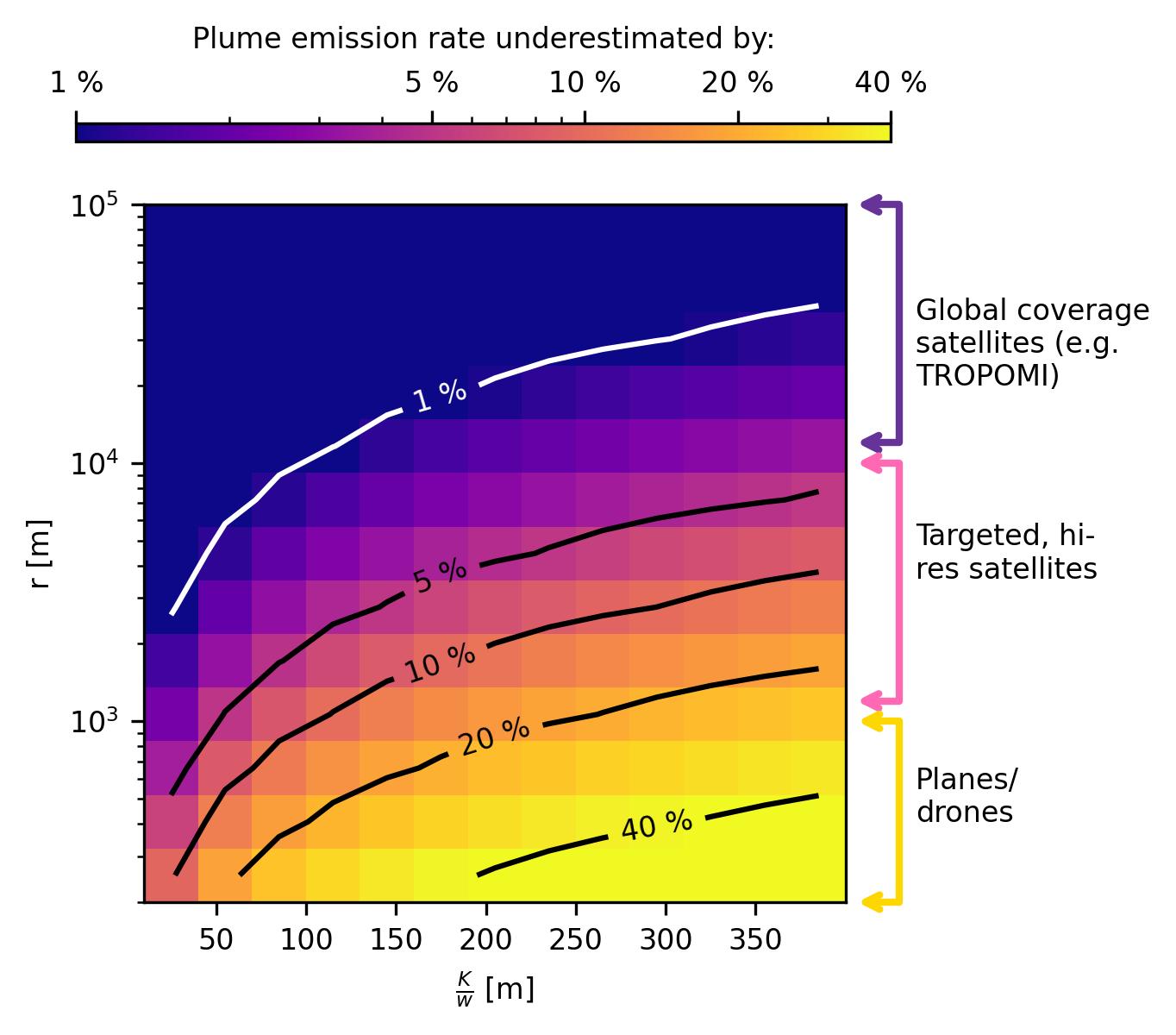}
\caption{A contour plot demonstrating the extent to which a plume's emission rate will be underestimated as a function of radius of integration $r$ and $K/w$, the ratio of the constant of turbulent diffusion to the wind speed. Shown on the right had side of the plot are general regions where certain observing methodologies would lie on this spatial scale. The range of values of $K/w$ (i.e., the $x$-axis) are chosen to represent a range of physically realistic values, estimated from high-resolution GHGSat scenes containing methane plumes. Global coverage satellites like TROPOMI tend to have larger pixel sizes and large fields of view, and hence are less likely to be affected by the negative bias. Targeted satellites can have narrower fields of view on the order of km scales, but due to their higher pixel resolutions can still image plumes on the scales of hundreds of meters. Consequently, these instruments may be affected by the negative bias of estimated emission rates, depending on how they are used. Surveys conducted via drone or plane may be especially susceptible to this bias if the divergence method is used to estimate emissions.}
\label{fig:4}
\end{figure}

The total area for spatial integration of the estimated emission field also influences $Q_{\mathrm{est}}$. In Fig. \ref{fig:3} we alter the radius $r$ of the cirular area of integration for a single simulated observation, and find that increasing the radius of integration increases $Q_{\mathrm{est}}$, i.e., the total estimated emission rate is less negatively biased. Although emissions are incorrectly distributed in the estimated emission field, they are distributed such that integrating the estimated emission field over a larger area improves the total estimated emission rate. Thus, we determine that in our procedure for estimating the emission rate of a plume, $r$ and $K/w$ are two independent parameters which determine the extent to which $Q_{\mathrm{est}}$ is underestimated. In Fig. \ref{fig:4} we vary $r$ and $K/w$ over physically realistic values (10 to 400 m$^2$ s$^{-1}$, see Sec. \ref{sec:permian_case_study} and \ref{sec:estimating_K} for a description of methods) and examine how greatly $Q_{\mathrm{est}}$ is underestimated for an ideal Gaussian plume resulting from a point source of emission. We find that in the most extreme cases, $Q_{\mathrm{est}}$ can be underestimated by more than 40\%. \\

In practice, different plume measurement methodologies will correspond to different parameter locations on Fig. \ref{fig:4}, and have characteristic biases associated with them.  We indicate three examples on the right hand edge of Fig. \ref{fig:4} of regions where certain instruments tend to lie. Global coverage satellites such as TROPOMI tend to have very large fields of view as well as lower pixel resolutions \cite{Veefkind_2012, Bovensmann_1999, Glumb_2014}. Regional emission budgets performed on the scale of tens of kilometers are unlikely to experience a high level of negatively biased emission estimates due to the lack of a diffusive term in the divergence method (e.g., see Sec. \ref{sec:permian_case_study}). More targeted satellites have higher pixel resolutions and are capable of imaging plumes on the scale of tens of meters \cite{Jervis_2021}. Although such satellites have fields of view that can still exceed ten kilometers or more, it would still be possible (given the high pixel resolution) to spatially integrate estimated emission fields over small enough areas to experience the outlined negative bias of the uncorrected divergence method, an issue that could arise if attempting to spatially isolate one plume from another adjacent source. Lastly, plume imaging surveys based from aircraft or drones \cite{Cusworth_2021, Ayasse_2022} would likely have the smallest field of view and would consequently be most likely to experience negative biases in emission estimates if they were to use the divergence method for emission rate estimation. \\

\begin{figure}[ht]
\centering
\includegraphics[width=7.2in]{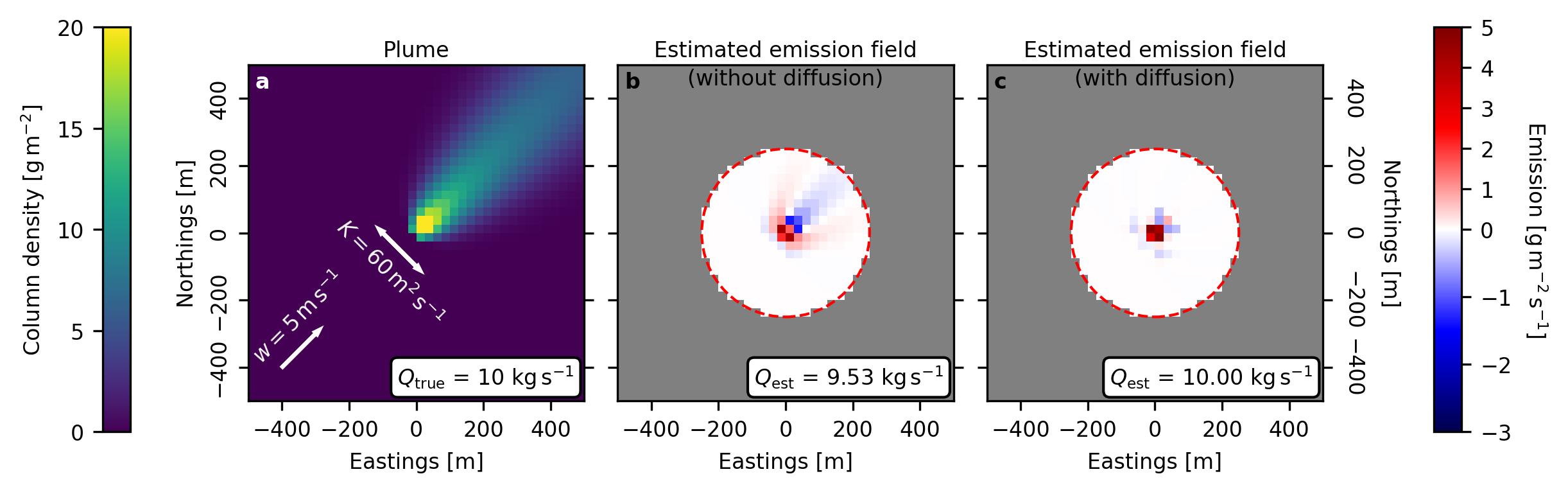}
\caption{A simulated column-integrated Gaussian plume, and resulting estimated emission fields via the divergence method. \textbf{a} A simulated observation of a plume. \textbf{b} The emission field of the plume estimated via the divergence method, using the emission equation $E = \nabla\cdot\vec{F}^\mathrm{adv}$. \textbf{c} The emission field of the plume estimated via the divergence method, using the emission equation $E = \nabla\cdot\left(\vec{F}^\mathrm{adv} - \vec{F}^\mathrm{dif}\right)$. The term $\vec{F}^\mathrm{dif}$ is calculated according to Eq. \ref{eq:F_dif_xy}. The emission field in \textbf{c} is now correctly estimated as a point source, with some small perturbations due to the numerical derivative. The correct total estimated emission rate $Q_\mathrm{est}$ is now obtained.}
\label{fig:5}
\end{figure}

The significance of diffusion on the accuracy of the divergence method can be seen most directly when a diffusive term is included in the emission equation, i.e., Eq. \ref{eq:corrected_emission_equation}. Making this addition to the method, we show in Fig. \ref{fig:5} that when the expression for diffusive flux (as calculated by Eq. \ref{eq:F_dif_xy}) is included in the emission equation, then the estimated emission field is correctly constrained to a point source (allowing some slight deviation due to numerical derivative effects). $Q_{\mathrm{est}}$ is found precisely to equal $Q_{\mathrm{true}}$. \\

A caveat to the efficacy of including diffusive flux in the divergence method to restore accurate emission estimates is that it is in practice difficult to estimate the constant of turbulent diffusion $K$. In our synthetic study, it is possible to know and choose the precisely correct value of $K$ to calculate $\vec{F}^\mathrm{dif}$, but for real data this is much harder to calculate. In Sec. \ref{sec:permian_case_study}, we estimate values of $K$ from high-resolution observations of real plumes.

\subsubsection{Miscalculating emission rates due to missing data}\label{sec:synth_study_missing_data}
Beirle et. al. (2019) \cite{Beirle_2019} was the first study to showcase the divergence method for estimating emission rates, and used TROPOMI observations of nitrogen dioxide to estimate emission rates from cities and power plants. They pointed out that, due to the linear properties of the divergence operator, it was sensible to time-average the daily fluxes of nitrogen dioxide first and then take the divergence of the time-averaged flux to obtain a time-averaged estimate of nitrogen dioxide emissions. Beirle et. al. (2019) outlined that this was a sensible procedure as the time-averaged nitrogen dioxide advective flux would have a smooth spatial distribution and thus allow for a more accurate calculation of spatial derivatives. \\

\begin{figure}[ht]
\centering
\includegraphics[width=7.2in]{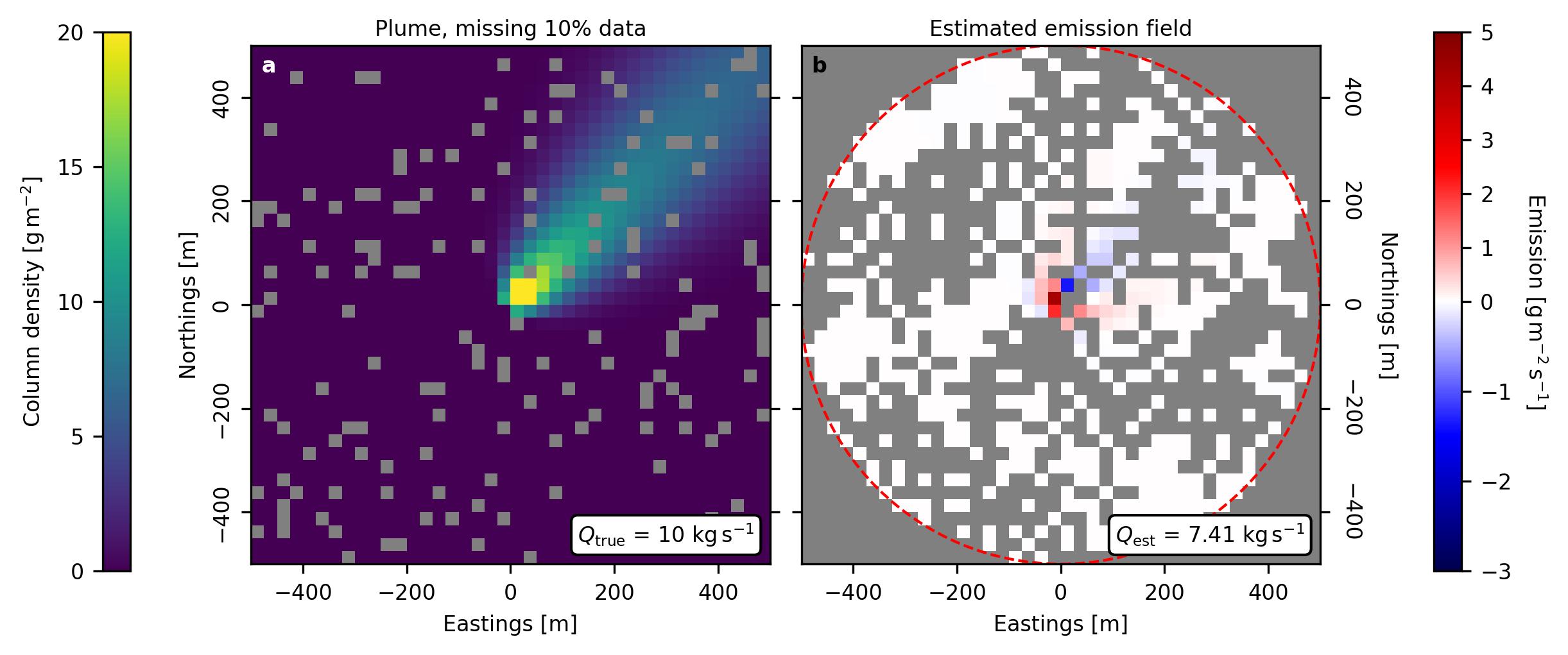}
\caption{\textbf{a} A simulated observation of a plume with wind angle $\theta = 45$ degrees, wind speed $w = 5\,\mathrm{m}\,\mathrm{s}^{-1}$, and $K=60\,\mathrm{m}^2\,\mathrm{s}^{-1}$. The point source of emission is located at the origin with actual emission rate $Q_\mathrm{true} = 10\,\mathrm{kg}\,\mathrm{s}^{-1}$. 10\% of the pixels in \textbf{a} are randomly masked to simulate the effect of poor data or otherwise unavailable observations, as is often the case for satellite observations of methane column density. \textbf{b} The resulting estimated emission field for the plume observation in \textbf{a}, obtained using the divergence method and emission equation $E = \nabla\cdot\vec{F}^\mathrm{adv}$. Our algorithm uses the fourth-order central finite difference to calculate numerical gradients \cite{Bronstein_1981}, which means that every unavailable pixel in \textbf{a} results in a 4x4 cross of pixels in \textbf{b} where we are unable to calculate $E$. Consequently, any amount of missing data in \textbf{a} quickly becomes a large amount of missing coverage in \textbf{b}. Integrating over the available pixels in \textbf{b} yields a severely underestimated estimated emission rate $Q_\mathrm{est}$. Scaling up $Q_\mathrm{est}$ by the fraction of missing coverage yields a scaled total emission estimate of 16.2 $\mathrm{kg}\,\mathrm{s}^{-1}$, and does not correct the underestimation. In Fig. S7 in the supplement we demonstrate that this scaling procedure is not sufficient to correct for the problem of missing data, and can actually significantly overestimate the total emission rate (as is the case in this particular example), depending on the fraction of daily missing data and number of repeated observations that go into the time-averaged calculation. In all other estimates of total emission rate in this study, we quote ``unscaled" results, i.e., the answer obtained by spatially integrating the estimated emission field over available grid cell values. It is also important to note that individual or time-averaged emission fields estimated via the divergence method in our synthetic study will vary depending on the spatial distribution of missing data, which is a random process. We do not cherry pick particular realisations of spatial distribution for missing data, and examine the variance that is inherent to this random process in Fig. \ref{fig:8}.}
\label{fig:6}
\end{figure}

TROPOMI observations of methane differ significantly from that of nitrogen dioxide, in that the spatial coverage of the methane data product is much less complete than the nitrogen dioxide data product \cite{NO2_ATBD, CH4_ATBD}. Fig. \ref{fig:6} demonstrates the problem this poses as we look to apply the divergence method to TROPOMI observations of methane; when we randomly mask 10\% of the observational data of a single simulated plume, the corresponding emission field estimated via the divergence method is missing a significantly larger amount of spatial coverage, and dramatically underestimates the total emission rate. This is because the numerical methods for calculating spatial derivatives \cite{Bronstein_1981, Gibou_2005} (see Eqs. \ref{eq:first_derivative} and \ref{eq:second_derivative}) require eight valid neighboring data points, and so any missing observational data is magnified into even more missing spatial coverage of the estimated emission field. Time-averaging observations prior to numerically calculating spatial derivatives, and choosing to describe average emission rates over time domains rather than at specific time points, hopefully allows a chance at circumnavigating the problem of truncated spatial observations.  \\

Thus, in this section we investigate two different methodologies for calculating time-averaged emission estimates using the divergence method. In the first methodology (which we denote by $\overline{E}_1$), we calculate daily estimates of emission fluxes and then temporally average them \cite{Veefkind_2023, Liu_2021}. Mathematically, this is represented via

\begin{equation}
    \overline{E}_1 = \overline{\nabla\cdot\vec{F}^\mathrm{adv}_t} \quad [\mathrm{kg}\,\mathrm{m}^{-2}\,\mathrm{s}^{-1}],
\end{equation}

\noindent where the subscript $t$ denotes advective fluxes on day $t$ in the time period of interest. In the second methodology (which we denote by $\overline{E}_2$), we temporally average the daily advective fluxes to create a time-averaged advective flux, and then take the divergence of the time-averaged advective flux to obtain a time-averaged estimated emission field \cite{Beirle_2019}. Mathematically, this is represented via 

\begin{equation}
    \overline{E}_2 = \nabla\cdot\overline{\vec{F}^\mathrm{adv}_t} \quad [\mathrm{kg}\,\mathrm{m}^{-2}\,\mathrm{s}^{-1}].
\end{equation}

\noindent  An explanation of these methodologies is given again briefly in Sec. \ref{sec:e1_e2} and in greater detail in supplementary section S4. In this section we do not correct the estimated emission fields of our simulated plumes for the effects of turbulent diffusion as we did in Sec. \ref{sec:synthetic_study_turbulent_diffusion}, and focus only on the differences between the $\overline{E}_1$ and $\overline{E}_2$ methodologies.\\

\begin{figure}[ht]
\centering
\includegraphics[width=7.2in]{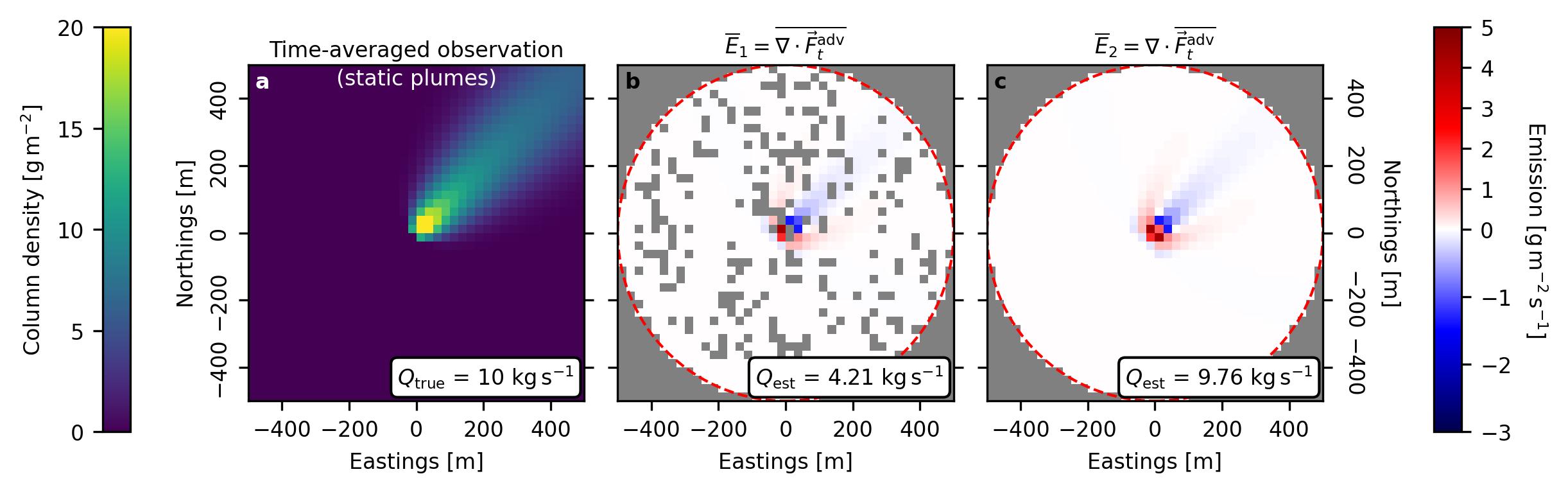}
\caption{\textbf{a} A time-averaged observation of 30 static plumes with constant plume parameters. Each individual plume has a random 30\% of its pixels masked (similarly to panel \textbf{a} in Fig. \ref{fig:6}). \textbf{b} The resultant average estimated emission field, constructed by time-averaging the estimated emission fields of each of the 30 individual plumes. \textbf{c} The average estimated emission field obtained by first time-averaging all the advective fluxes of the 30 individual plumes, and then taking the divergence of the time-averaged fluxes. The estimated emission field in \textbf{c} is much more resilient to the missing data, and provides a much more accurate estimate of the average emission rate than in \textbf{b}. }
\label{fig:7}
\end{figure}

In Fig. \ref{fig:7} we simulate a time-averaged study of 30 steady-state Gaussian plumes. Plume parameters are left unvaried, and each of the 30 repeated simulated observations has a random 30\% of its pixels masked. We then display the resulting estimated emission fields obtained via $\overline{E}_1$ and  $\overline{E}_2$, and find that the emission field of $\overline{E}_2$ is spatially complete, whereas $\overline{E}_1$ is still missing some spatial coverage. The total integrated emission rate of $\overline{E}_1$ is also severely underestimated, but the total integrated emission rate of $\overline{E}_2$ retrieves the correct time-averaged emission rate (apart from the slight negative bias due to the presence of diffusion in the plume, which was discussed in the previous section). Figures showing the difference in resulting estimated emission fields under non-static conditions are shown in the supplement in Figs. S2, S3, and S4. \\

\begin{figure}[ht]
\centering
\includegraphics[width=7.2in]{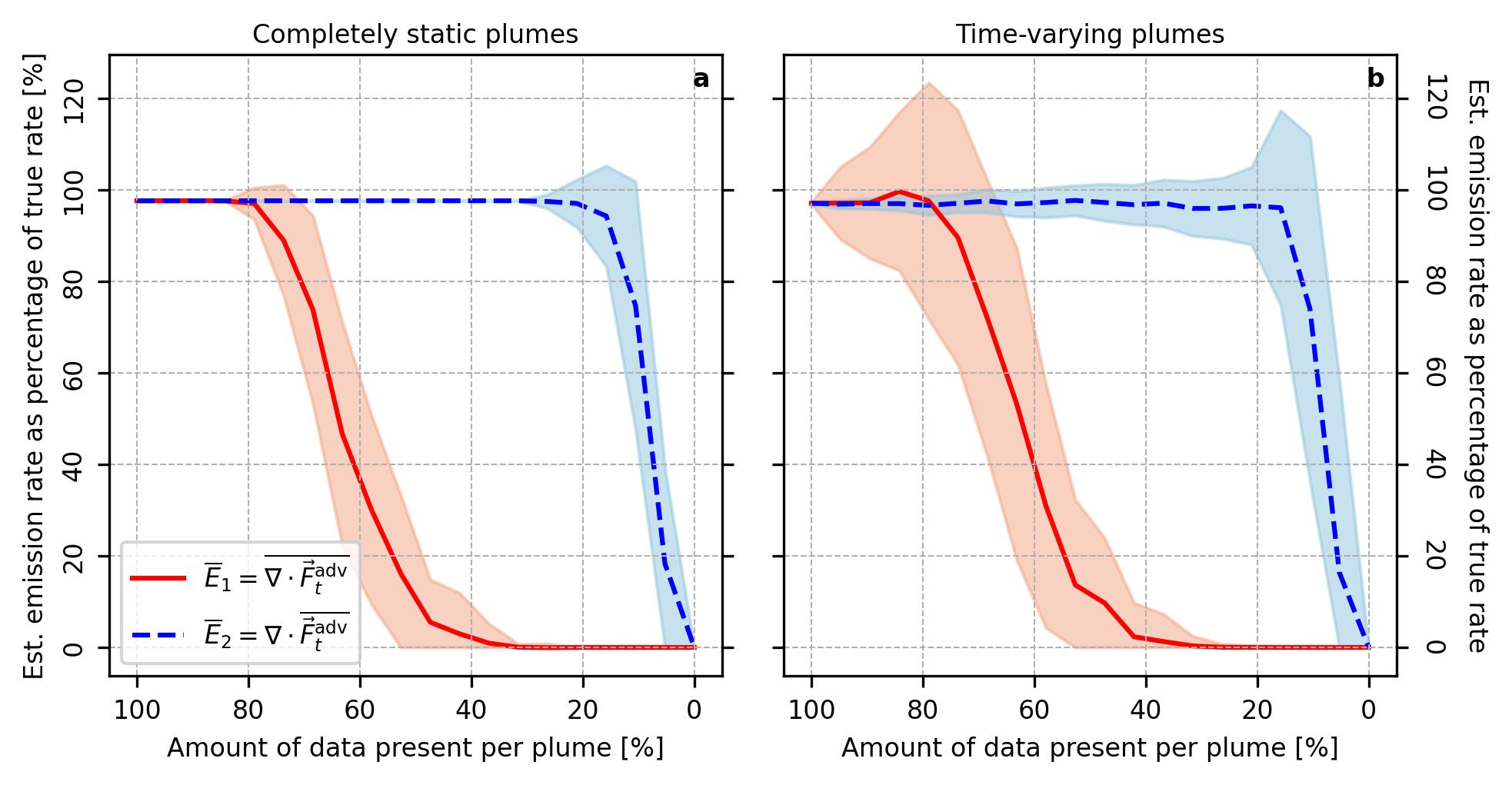}
\caption{Underestimation of time-averaged estimated emission rate, plotted as a function of the percentage of observational data present for each individual plume that goes into into the time-averaged estimation. In \textbf{a}, each time-averaged calculation uses 30 individual plumes, each one static with constant wind speed, angle, point source location and emission rate, etc. Each of the 30 simulated observations differs only in the spatial distribution of the masked pixels. In \textbf{b}, each time-averaged calculated uses 30 individual plumes, but the wind speed, wind angle, and source location are randomised. Each of the 30 simulated observations then have the same percentage of pixels randomly masked. This scenario represents a higher degree of complexity, where time-invariant assumptions about sources of emission no longer hold. We find that for both scenarios, it is better to time-average advective fluxes and take the divergence once, as opposed to time-averaging individually estimated emission fields (from the perspective of accurate emission flux estimation).}
\label{fig:8}
\end{figure}

We investigate how the difference in performance of the $\overline{E}_1$ and $\overline{E}_2$ methodologies varies as a function of amount of daily missing data, and when plume parameters are allowed to vary in time. These results are shown in Fig. \ref{fig:8}. We find that as the amount of missing observational data increases, both methodologies underestimate the true average emission rate, but that method $\overline{E}_2$ (i.e., averaging daily fluxes of $C$ and taking the divergence once) allows for estimates that are far more robust against missing data. This holds true for both static and time-varying simulated plume observations, although the time-averaged estimated emission rates for the time-varying plumes have more variance than the static plumes. This simulation differs from realistic physical scenarios in that data is randomly masked in a physically uncorrelated manner (which would not be the case with cloud cover), but it nonetheless demonstrates that the way in which time-averaged emission estimates are calculated using the divergence method is not trivial. Additionally, Fig. \ref{fig:8} demonstrates that it is also possible to over-estimate the source emission rate, though this typically only occurs at a critical ``turn over" point where the fraction of daily missing data begins to dominate over the number of repeated observations. Past this point, we find that estimated emission rates will only be underestimated as a consequence of spatially incomplete data, but the exact location of this critical value is highly dependant on the number of repeated observations and the spatial distribution of missing data.

\subsection{Permian basin case study}\label{sec:permian_case_study}
\begin{table}
\centering
\caption{\label{tab:previous_work}Estimates of Permian methane emission rates [Tg/year]. We compare our yearly estimates via the $\overline{E}_2$ methodology (where daily methane fluxes are averaged) to those in other literature and find good agreement. Uncertainties on our time-averaged emission estimates are calculated via the algebraic propagation of the daily variance of advective flux of methane at each grid cell. This methodology is described in supplementary section S4.}
\begin{threeparttable}
\begin{tabular}{@{}llllll}
\hline \\ \vspace{-2em} \\
Year & This work & Veefkind et. al. 2023$^*$ &  Schneising et. al. 2020$^{||}$ & Liu et. al. 2021  & Zhang et. al. 2020$^+$ \\
2019 & 3.1 $\pm$ 0.7 & 3.0 & 2.9 $\pm$ 1.6 & 3.1 (2.8, 3.8) & 2.7 $\pm$ 0.5 \\
2020 & 2.4 $\pm$ 0.6 & 2.8 & 2.3 $\pm$ 1.7 & - & - \\
2021 & 2.7 $\pm$ 0.5  & - & - & - & - \\
\vspace{-1em} \\
\hline
\end{tabular}
\begin{tablenotes}\footnotesize
\item[] $^*$Calculated using TROPOMI methane mixing ratios from the Weighting Function Modified Differential Optical Absorption Spectroscopy (WFM-DOAS) algorithm version 1.5 \cite{Schneising_2019}. 1 - $\sigma$ uncertainty of 25\%.
\item[] $^+$Date range for this paper was March 2018 - March 2019.
\item[] $^{||}$Value taken from Veefkind et. al. 2023 \cite{Veefkind_2023}, i.e., calculated using the methods of Schneising et. al. 2020 \cite{Schneising_2020}, but using the same data as Veefkind 2023 \cite{Veefkind_2023}.
\end{tablenotes}
\end{threeparttable}
\end{table}

\begin{table}
\centering
\caption{\label{tab:method_comparison}Comparison of yearly methane emission rate estimates [Tg/year] for the Permian when using the two different time-averaging methodologies. In the first column we show yearly methane emission rate estimates for the Permian via $\overline{E}_2$, when we average over advective fluxes for the year and take the divergence to yield a time-averaged emission estimate. In the next column we show the same yearly emission rate estimate calculated via $\overline{E}_1$, where we calculate daily methane emission estimates and time-average them. In the penultimate column we show the difference of the total estimated emission rate between the two methodologies, when the estimated emission fields are only spatially integrated over the intersection of the two estimated emission fields. This is to examine whether the difference in the estimated emission rate is driven by differences in spatial coverage or not. Supplementary figures 5, 6, and 7 plot these estimated emission fields. Also shown in the last column is the average daily spatial coverage of the Permian basin by our regridded TROPOMI methane observations.}
\begin{threeparttable}
\begin{tabular}{@{}lllll}
\hline \\ \vspace{-2em} \\
 Year &  $\overline{E}_2 = \nabla\cdot\overline{\vec{F}^\mathrm{adv}_t}$ &  $\overline{E}_1 = \overline{\nabla\cdot\vec{F}^\mathrm{adv}_t}$ & $\cap\left(\overline{E}_2-\overline{E}_1\right)$ & Avg. Daily Coverage$^*$\\
2019 & 3.06 $\pm$ 0.66 & 1.45 $\pm$ 0.31 & 0.31 $\pm$ 0.57 & 29.37 \%\\
2020 & 2.39 $\pm$ 0.55 & 1.25 $\pm$ 0.30 & 1.16 $\pm$ 0.62 & 37.68 \%\\
2021 & 2.67 $\pm$ 0.54 & 1.75 $\pm$ 0.26 & 0.92 $\pm$ 0.60 & 41.15 \%\\
\vspace{-1em} \\
\hline
\end{tabular}
\begin{tablenotes}\footnotesize
\item[] $^*$This is average percentage coverage of the Permian basin by the Copernicus Sentinel-5P TROPOMI Level 2 methane data product for each day within the given year. This is not the percentage coverage of the Permian basin by the time-averaged estimated emission field for the year.
\end{tablenotes}
\end{threeparttable}
\end{table}

The Permian basin is the largest oil and gas producing region in the United States, producing nearly 6,000 barrels of oil a day as of January 2023 \cite{Permian_2023}. Due to its prominence and size the Permian is frequently a target of ground-based, airborne, and space-based campaigns monitoring methane emissions \cite{Gouw_2020, Robertson_2020}. We grid three years of daily TROPOMI methane observations of the Permian basin (2019-2021) onto a 0.2 x 0.2 latitude-longitude grid using an area-weighted oversampling \cite{Zhu_2017} and calculate yearly emission estimates. \\

\begin{figure}[ht]
\centering
\includegraphics[width=3.5in]{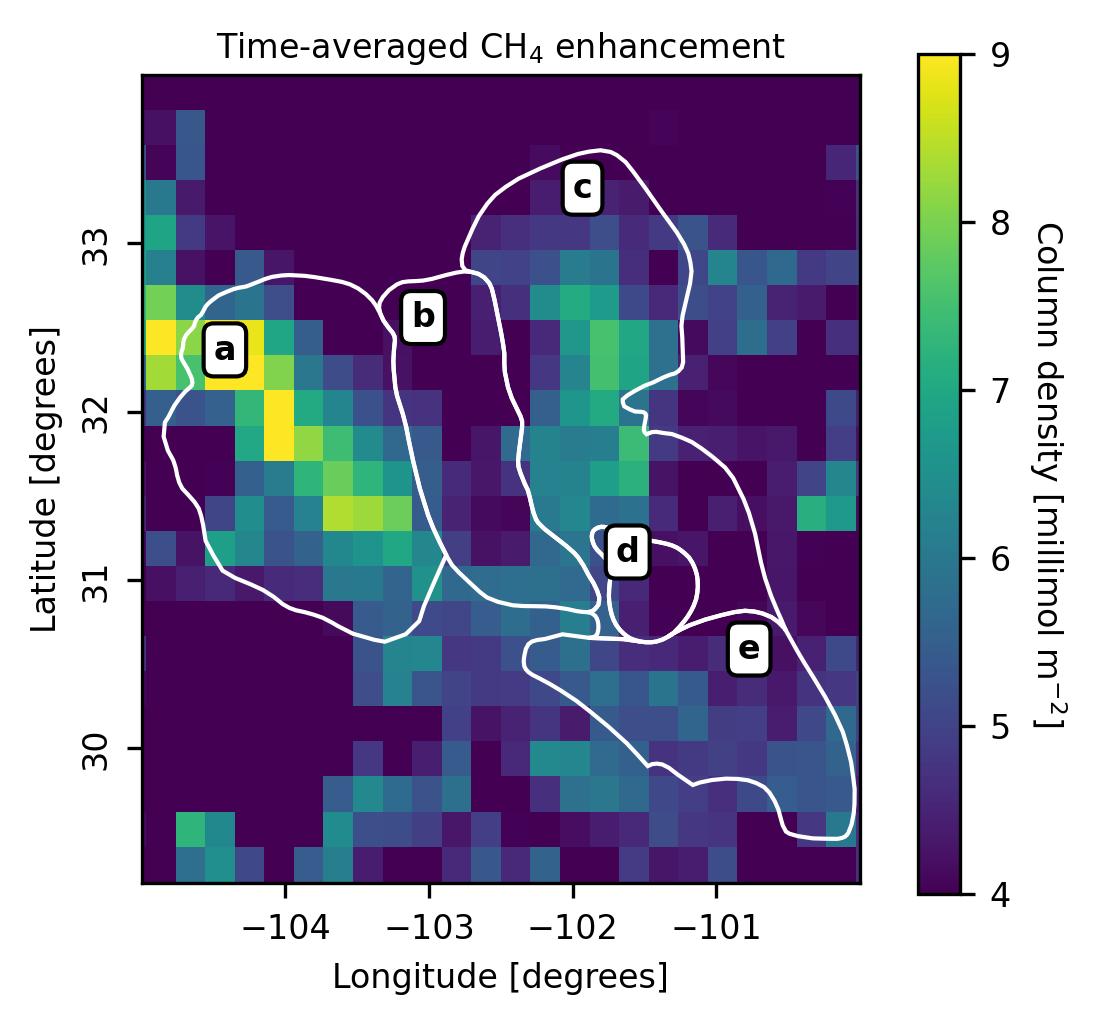}
\caption{Average methane enhancement over the Permian basin for the years 2019, 2020, and 2021 as observed by TROPOMI. \textbf{a} The Delaware basin. \textbf{b} The Central basin. \textbf{c} The Midland basin. \textbf{d} The Ozana Arch basin. \textbf{e} The Val Verda basin. Basin boundaries taken from the Energy Information Administration.}
\label{fig:9}
\end{figure}

We use the Copernicus Sentinel-5P TROPOMI Level 2 methane data product \cite{TROPOMI_CH4_DATA}, which provides daily global methane observations as total column-averaged methane mixing ratios X(CH$_4$). We convert X(CH$_4$) [ppbv] to total column densities $n$ [mol m$^{-2}$] via

\begin{equation}
    n = 10^{-9}\,\mathrm{X(CH}_4\mathrm{)}\,\frac{p}{g\,m_\mathrm{air}} \quad [\mathrm{mol}\,\mathrm{m}^{-2}],
\end{equation}

\noindent where $p$ is the dry surface pressure in Pa, $g$ is the constant of gravity (9.81 m s$^{-2}$), and $m_\mathrm{air}$ is the molar mass of dry air ($28.96\times 10^{-3}$ kg mol$^{-1}$) \cite{Veefkind_2023}. We determine dry surface pressure from the ERA5 reanalysis dataset from the European Centre for Medium-Range Weather Forecasts (ECMWF) \cite{Hersbach_2020}. The total column density $n$ must further be background-reduced before applying the divergence method. Following the methodology of Veefkend et. al. (2023), methane column enhancements $\Delta n$ above background can be determined on a daily basis via

\begin{equation}
    \Delta n = n - \left(c_0 + c_1\,p\right) \quad [\mathrm{mol}\,\mathrm{m}^{-2}],
\end{equation}

\noindent where $c_0$ and $c_1$ are constants determined from a linear fit of $n$ vs $p$. This methodology assumes a limited-size area of interest, such as the Permian, and further assumes that the tropopause pressure is constant over the area of interest \cite{Veefkind_2023}. The 2019-2021 average methane enhancement over the Permian basin is shown in Fig. \ref{fig:9}.  \\

\begin{figure}[ht]
\centering
\includegraphics[width=7.2in]{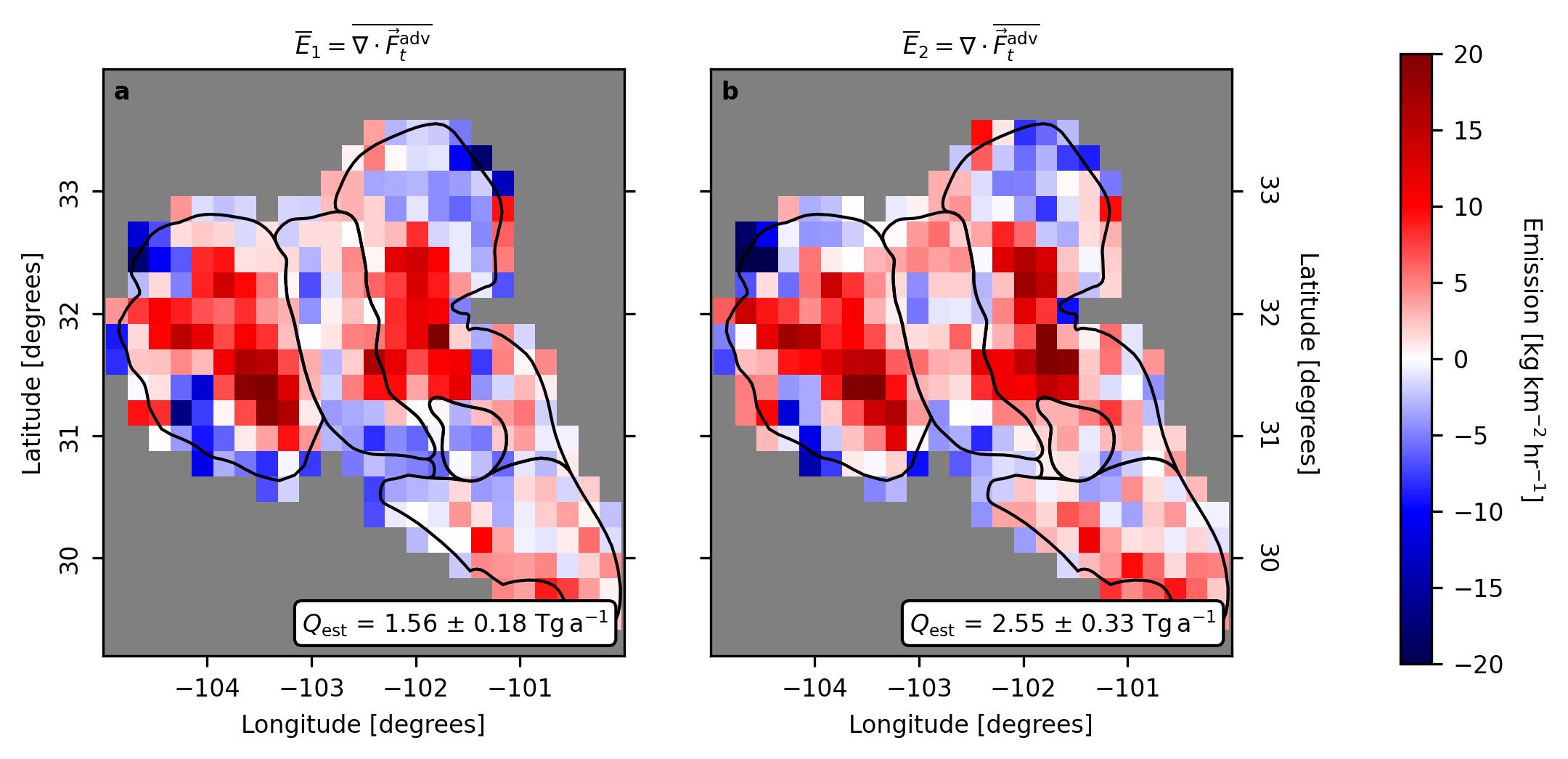}
\caption{Estimated emission fields for the five Permian subbasins for three years of observations, 2019-2021. \textbf{a} Estimated emission field obtained when daily estimated emission fields are averaged. \textbf{b} Estimated emission field when daily advective fluxes are averaged, with the divergence taken thereof to obtain the estimated emission field. Although the extended three-year time period means that complete spatial coverage is achieved for all the basins, the intermittent missing data on a daily basis means that the methodology of $\overline{E}_1$ estimates a significantly lower emission rate than the methodology of $\overline{E}_2$. We find the three-year estimated emission fields $\overline{E}_1$ and $\overline{E}_2$ to be correlated with $r=0.87$ . }
\label{fig:10}
\end{figure}

Using ERA5 wind data from the ECMWF \cite{Hersbach_2020}, we calculate daily advective fluxes of methane, and then calculate yearly time-averaged methane emission maps of the Permian basin using both the $\overline{E}_1$ and $\overline{E}_2$ methodology. As a reminder: in the $\overline{E}_1$ methodology, we use the divergence method to estimate daily emission fields, which are then time-averaged, and in the $\overline{E}_2$ methodology, we first time-average daily advective fluxes methane, and use the divergence method to estimate the emission field from the time-averaged fluxes. As in Sec. \ref{sec:synth_study_missing_data}, for simplicity $\overline{E}_1$ and $\overline{E}_2$ both exclude the turbulent diffusive flux term, in order to purely examine the differences arising from the choice of how to order the time-averaging operations (though later in this section, we examine the extent to which turbulent diffusion matters for this observing system). The yearly estimated emission fields produced via the two methodologies are shown in the supplement, and the estimated emission fields produced via the two methodologies for the entire time period 2019-2021 is shown in Fig. \ref{fig:10}. Our yearly total estimated methane emission rates for the Permian are shown in Tables \ref{tab:previous_work} and \ref{tab:method_comparison}. We find good agreement between our time averaging methodology $\overline{E}_2$ and other Permian emission estimates from previous work, but find that the $\overline{E}_1$ methodology (in which daily emission estimates for the Permian are time-averaged) significantly underestimates the time-averaged emission rates when compared to previous estimates in the literature. It should be noted that some of the results of previous work quoted in Table \ref{tab:previous_work} use TROPOMI methane columns calculated via the Weighting Function Modified Differential Optical Absorption Spectroscopy
(WFM-DOAS) algorithm version 1.5 \cite{Schneising_2019}, an alternative data product which has a greater spatial coverage than the traditional Copernicus Sentinel-5P TROPOMI Level 2 methane data product. Consequently, these results are less likely to be susceptible to the weaknesses we note here in using the divergence method with spatially corrupted data. \\

The difference in results between our $\overline{E}_1$ and $\overline{E}_2$ methodologies is likely due to the sparse daily spatial coverage of the TROPOMI methane data product over the Permian basin. For the years 2019-2021, the average daily coverage of our regridded TROPOMI observations of the Permian basin never exceeds 50\% (Table \ref{tab:method_comparison}). In Table \ref{tab:order_choice}, we examine whether the choice of order of central finite difference influences the results obtained when calculating time-averaged emission rate estimates for the Permian. Although the spatial coverage of the estimated emission field produced via $\overline{E}_1$ improved by using the second order central finite difference to calculate derivatives instead of the fourth order central finite difference, we did not find any significant change in the total estimated emission rates.\\ 

\begin{table}
\caption{Yearly estimates of methane emission from the Permian basin [Tg/year]. We present yearly estimates (calculated using both the $\overline{E}_1$ and $\overline{E}_2$ time-averaging methodologies), and compare results between when using the fourth-order central finite difference and the second-order central finite difference to calculate numerical derivatives. The second-order central finite-difference requires fewer valid neighbors to calculate derivatives, and so could potentially lessen the discrepancy between the results of the $\overline{E}_1$ and $\overline{E}_2$ methodologies. We find that for the year 2019 (which had the poorest average spatial coverage over the Permian by the TROPOMI methane data product), the difference between the yearly methane emission budgets estimated via $\overline{E}_1$ and $\overline{E}_2$ is slightly decreased, but the gap between the two is not bridged within error. Also shown in this table is the percentage area coverage of the Permian basin by the estimated emission field produced by the $\overline{E}_1$ methodology. As expected, the percentage area coverage is improved by the usage of the second order central finite difference in calculating derivatives, and the greatest improvement is seen in the year 2019. However, this increase in spatial coverage of the estimated emission field is not sufficient alone in bridging the discrepancy of results produced by the $\overline{E}_1$ and $\overline{E}_2$ methodologies. }
\centering
\label{tab:order_choice}
\begin{threeparttable}
\begin{tabular}{@{\extracolsep{8pt}}l c l c c l c @{}}
\toprule
\noalign{\vspace{0.5em}}
\multirow{2}{*}{Year} & \multicolumn{3}{c}{4th order CFD} & \multicolumn{3}{c}{2nd order CFD} \\ 
\noalign{\vspace{0.5em}}
\cline{2-4}\cline{5-7}
\noalign{\vspace{0.5em}}
  & $\overline{E}_1$ & $\overline{E}_1$ \% coverage$^*$ & $\overline{E}_2$ & $\overline{E}_1$ & $\overline{E}_1$ \% coverage$^*$ & $\overline{E}_2$  \\ 
\noalign{\vspace{0.25em}}
\cline{1-4}\cline{5-7}
\noalign{\vspace{0.5em}}
2019 & 1.45 $\pm$ 0.31 & 73.29\% & 3.06 $\pm$ 0.66 & 1.84 $\pm$ 0.26 & 89.53\%  & 3.10 $\pm$ 0.49  \\

2020 & 1.25 $\pm$ 0.30 & 98.19\% & 2.39 $\pm$ 0.55 & 1.20 $\pm$ 0.22 & 100\%  & 2.37 $\pm$ 0.41 \\

2021 & 1.75 $\pm$ 0.26 & 98.92\% & 2.67 $\pm$ 0.54 & 1.63 $\pm$ 0.20 & 99.64\% & 2.61 $\pm$ 0.41 \\

\bottomrule
\end{tabular}
\begin{tablenotes}\footnotesize
\item[] $^*$This is the percentage coverage of the Permian basin of the time-averaged estimated emission field for this year, and not the average daily coverage of the Permian basin for the year by the methane data product.
\end{tablenotes}
\end{threeparttable}
\end{table}

As the total estimated emission rate for the Permian is obtained via spatially integrating the estimated emission field over the extent of the Permian, it follows that a methodology which produces greater spatial coverage ($\overline{E}_2$) will produce a larger total estimated emission rate than a method with less spatial coverage ($\overline{E}_1$), assuming that grid cells are generally estimated to have positive emissions. However, in Table \ref{tab:method_comparison} we show the difference in total estimated emission rate between $\overline{E}_1$ and $\overline{E}_2$ for when we only spatially integrate over the intersection of their spatial coverages. In this case, any difference in total estimated emission rate for the overlapping coverage is due to the differing methodologies of $\overline{E}_1$ and $\overline{E}_2$, rather than the difference in spatial coverage of the Permian they produce. We find that for the year 2019, the difference in average total estimated emission rate between $\overline{E}_1$ and $\overline{E}_2$ can potentially be explained entirely by the fact that the estimated emission field produced via $\overline{E}_2$ covers a significantly larger geographic area than that produced via $\overline{E}_1$. This is shown in Fig. S4. However, for the years 2020 and 2021, the spatial coverage of the estimated emission fields obtained via $\overline{E}_1$ and $\overline{E}_2$ are complete, and thus the difference in total estimated emission rates cannot be explained by a difference in spatial coverage. Increasing the spatial coverage of the region of interest by the methane data \cite{Roberts_2022, Schneising_2019} may close the gap in the results obtained between the two methods. This indicates that an analytical determination of the constant of turbulent diffusion $K$ from observations may only be suitable for large, plume-like emitters and would not be applicable to large-scale diffuse emission environments.\\

We additionally calculate the percent change of our yearly estimated methane rates for the Permian when including the diffusive flux calculation of Eq. \ref{eq:cartesian_F_dif}. Based on a number of real methane plumes measured by GHGSat \cite{Varon_2019}, a typical value of $K$ was determined to be highly variable and in the the range between 10 and 400 $\mathrm{m}^2\,\mathrm{s}^{-1}$ (see Sec. \ref{sec:estimating_K} for a description of methods). We find in all three year-averaged emission estimates for the Permian that the total estimated emission rate is increased by less than a millionth of a percent when $K=400\,\mathrm{m}^2\,\mathrm{s}^{-1}$, the maximum value for the constant of turbulent diffusion that we consider in this work. Furthermore, the inclusion of the turbulent diffusion flux term (even at this strength) does not change the spatial distribution of estimated emissions, and still shows emission enhancements over the Midland and Delaware basins. This is not unexpected given the results shown in Fig. \ref{fig:4}, which suggests that integrating over large areas sufficiently corrects for any negative bias introduced by neglecting turbulent diffusion in the divergence method. However, the plume observations \cite{Varon_2019} used to calculate the potential range of values of $K$ may not be indicative of typical plume morphology for other geographic locations or meteorological conditions. If satellite observations are not screened for low wind fields, greater emission underestimates may be obtained via the divergence method. Smaller plumes than those shown in this example study may exhibit different behaviour, though they may not be resolvable on TROPOMI-pixel scales.

\section{Discussion}
In this work, we examine the conditions under which the divergence method for estimating emission rates may prove to produce negatively biased results. Using a simulation study with synthetic satellite observations of ideal Gaussian plumes, we showed that highly-resolved, diffuse plumes may have negatively biased emission rates when their estimated emission fields are spatially integrated over narrow fields of view. Our simulation study suggests that this affect would only be of concern for observations obtained by high-resolution, narrow field of view methodologies, e.g., drones, or where high-resolution satellite data has clipped the area into which a plume extends. In contrast, our case study of the Permian basin with TROPOMI methane observations does not find that yearly estimated methane emission budgets are impacted by including even a high estimate of turbulent diffusion. In the future, as satellites become more capable of resolving individual plumes, it will become important to correct for turbulent diffusion when estimating the spatial distribution of emissions via the divergence method. In these scenarios where fine-scale plume structure can be spatially resolved, emission rate estimation is likely better performed via the IME or plume rotation methods, as the divergence method (like the inverse Gaussian plume method) is unable to adequately take turbulence into account \cite{Varon_2018}. Conditions where diffusion may be dominating over advection may also be identified by screening for very low wind speeds \cite{Veefkind_2023}. \\

In our synthetic study, we do not model the effects of instrumentation error or averaging kernel sensitivities \cite{eskes_averaging_2003}; this is in order to isolate and examine the effects of missing data and the turbulent diffusion flux term. We do give an example supplementary figure (Fig. S1) showing a simulated plume observation and resulting estimated emission field where some added noise is included in the observation. Previous work \cite{Varon_2018} also shows that the Gaussian plume model does not accurately model the fine-scale turbulent eddy structure for highly resolved plumes; although this means our synthetic plume observations are more idealised than real plumes, it allows us to test the assumptions of the divergence method systematically against a simple test dataset in a robust manner. We do not mean to suggest that the inclusion of the turbulent diffusion flux term in the divergence equation would entirely correct the calculation for real plumes, but that it is likely a necessary first-order correction.\\

We also examine two possible methodologies for calculating time-averaged emission estimates using the divergence method. Using simulated spatially incomplete plume observations, we find that time-averaging daily emission rate estimates produced via the divergence method will consistently underestimate the true average emission rate. We compare these results to an alternative methodology previously described for nitrogen dioxide observations \cite{Balcombe_2018}. By this method, daily advective fluxes are time-averaged, and the divergence is taken thereof in order to obtain a time-averaged emission estimate. We find in our synthetic study that this latter methodology yields robust emission estimates even in the face of spatially incomplete observations. We compare these two methodologies by constructing yearly methane emission budget estimates for the Permian basin for the years 2019-2021, using the Copernicus Sentinel-5P TROPOMI Level 2 methane data product \cite{TROPOMI_CH4_DATA}. We find that these two methodologies do not produce congruent emission rate estimates, and that the latter methodology produces estimates in agreement with previous top-down estimates for methane emission in the Permian basin \cite{Zhang_2020, Veefkind_2023}. Methods and datasets exist that can augment the spatial coverage of the TROPOMI methane data product \cite{Schneising_2019, Roberts_2022}, which in turn would augment the spatial coverage of the advective flux field of methane prior to the application of the divergence method. Spatial smoothing and interpolation could also be used to try and make the spatial coverage of the advective methane flux field more complete. In this work, we do not explore these avenues further, preferring to examine the differences between the $\overline{E}_1$ and $\overline{E}_2$ methodologies.  \\

When using the divergence method for methane emission rate estimation in the Permian, we find areas bordering the Delaware and Midland basins that are estimated to have negative emission rates. We do not expect this to truly be the case. Even in the ``perfect" case in our synthetic study when the point source of emission is correctly estimated (see Fig. \ref{fig:5}, panel \textbf{c}), we still estimate some grid cells to have negative emission rates. In this case, this is an effect of the discretisation of the emission equations and the manner in which numerical derivatives are calculated over our grids. Therefore, in some areas, especially in the vicinity of large methane sources, the divergence method will return negative emissions. These are local artefacts; the area-integrated emissions remain positive both in our synthetic study and our case study of the Permian basin (and in this latter case, our results are in good agreement with previously estimated methane emission rates). Whilst the TROPOMI methane observations over the Permian can not be considered to be ideal plumes, it may be the case that the regions of negative estimated emission are analogous to those in the synthetic study, as we know that that they are bordering known regions of strong positive emission. Other work also demonstrates that some regions of negative emission estimated via the divergence method in the Permian can be related to changes in orogoraphy, surface albedo, or convergent wind fields \cite{Veefkind_2023}. Other works demonstrate that time-averaged emission calculations from the divergence method are unlikely to be dominated by convergent wind fields \cite{Liu_2021}, and other formulations of the divergence method calculate advective flux without any contribution from wind field divergence \cite{beirle_improved_2023, sun_derivation_2022}. One could develop a model that prohibits the estimation of negative methane emissions (other works do so in a Bayesian framework \cite{Zhang_2020, maasakkers_global_2019}), though at this stage this would no longer purely be the ``divergence method", which is driven entirely by the data and the principle of the conservation of mass. \\

We conclude that the divergence method for estimating methane emissions (as described in this work) would be best applied to regional analyses where the affects of turbulent diffusion are unlikely to dominate over the scale of advective methane fluxes. Under these scenarios, we have demonstrated that it is possible to obtain accurate total estimated emission rates due to the large area of spatial integration. For such larger-scale analyses, it is important to be able to accurately determine the spatial distribution of emissions in order to identifying regions of high emissions, which may not always align with reported bottom-up inventories. In contrast, emission quantification for point sources with known locations (e.g., facility-scale emission audits) may be better achieved using alternative methodologies such as IME or the plume rotation method. When planning emission measurement campaigns, it is important to identify whether the primary goal is emission location or quantification, as different methodologies may prove better suited to one goal or the other. Whenever possible, spatially completely methane data products should be used. When using spatially incomplete datasets, it may be the case that taking the divergence of time-averaged advective fluxes of methane will produce more accurate methane emission rate estimates. 

\section{Methods and Data}
\subsection{Generating simulated satellite observations of plumes}
For our synthetic studies we generate simulated top-down observations of ideal Gaussian plumes \cite{Stockie_2011} via the equation

\begin{equation}\label{eq:plume_generation_equation}
    C\left(x,\,y,\,\theta\right) = \frac{Q}{2\sqrt{\pi\,w\,K\,\left(x\,\mathrm{cos}\theta + y\,\mathrm{sin}\theta\right)}}\mathrm{exp}\left[-\frac{\left(y\,\mathrm{cos}\theta - x\,\mathrm{sin}\theta\right)^2\,w}{4\,K\,\left(x\,\mathrm{cos}\theta + y\,\mathrm{sin}\theta\right)}\right]\quad\left[\mathrm{kg}\,\mathrm{m}^{-2}\right],
\end{equation}

\noindent where $Q$ is the point source emission rate [$\mathrm{kg}\,\mathrm{s}$], $w$ is the wind speed [$\mathrm{m}\,\mathrm{s}^{-1}$], $K$ is the constant of turbulent diffusion [$\mathrm{m}^2\,\mathrm{s}^{-1}$], and $\theta$ is the wind angle relative to the $x$-axis (in the anti-clockwise direction). Eq. \ref{eq:plume_generation_equation} is derived in S1 . There are a variety of assumptions that are fundamental to the ideal Guassian plume equation \cite{Stockie_2011}, but most important of note here is that diffusion is assumed to be dominated by advection, and thus diffusion only takes place perpendicular to the wind vector characterised by $w$ and $\theta$.

\subsection{Calculating emissions and flux terms}
We calculate spatially varying estimated emission fields $E$ using both simulated plume observations and the TROPOMI L2 methane data product. We calculate $E$ via two different emission equations. \\

The first emission equation (commonly found in literature \cite{Beirle_2019, Veefkind_2023}) is 

\begin{equation}\label{eq:simple_emission_equation}
    E = \nabla\cdot\vec{F}^\mathrm{adv} \quad \left[\mathrm{kg}\,\mathrm{m}^{-2}\,\mathrm{s}^{-1}\right],
\end{equation}

\noindent where $\vec{F}^\mathrm{adv}$ [$\mathrm{kg}\,\mathrm{m}^{-1}\,\mathrm{s}^{-2}$] is the advective flux of some column density $C$. In the case of our synthetic plume observations, $C$ is generated via Eq. \ref{eq:plume_generation_equation}. For TROPOMI observations of methane, we convert column-averaged mixing ratios to above-background column density enhancements \cite{Veefkind_2023}. $\vec{F}^\mathrm{adv}$ is then given by 

\begin{equation}\label{eq:F_adv_xy}
    \vec{F}^\mathrm{adv} = C\,\vec{w} \quad[\mathrm{kg}\,\mathrm{m}^{-1}\,\mathrm{s}^{-1}],
\end{equation}

\noindent where $\vec{w}$ is a spatially varying wind vector with magnitude $w$ and angle $\theta$ relative to the $x$-axis of our grid. For our synthetic studies we specify $w$ and $\theta$ ourselves. For our work with TROPOMI observations of the Permian basin, we take $\vec{w}$ to be the ERA5 wind data on multiple pressure levels, temporally averaged daily over a wind history at 1700, 1800 and 1900 hours, and then averaged vertically to an altitude of 500m to account for changes in wind vector through the boundary layer \cite{Veefkind_2023}. \\

To examine the extent to which turbulent diffusion influences estimated emission rates via the divergence method, we also calculate $E$ via a second emission equation 

\begin{equation}\label{eq:corrected_emission_equation}
    E = \nabla\cdot\left(\vec{F}^\mathrm{adv} - \vec{F}^\mathrm{dif}\right) \quad \left[\mathrm{kg}\,\mathrm{m}^{-2}\,\mathrm{s}^{-1}\right]. 
\end{equation}

\noindent $\vec{F}^\mathrm{dif}$ is the turbulent diffusive flux of some column density $C$, and for an ideal Guassian plume is given by 

\begin{equation} \label{eq:F_dif_xy}
    \vec{F}^\mathrm{dif} = K\,\left(\frac{\partial\,C}{\partial\,x}\,\mathrm{sin}^2\theta - \frac{\partial\,C}{\partial\,y}\,\mathrm{cos}\theta\,\mathrm{sin}\theta\right)\,\vec{e_x} + K\,\left(\frac{\partial\,C}{\partial\,y}\,\mathrm{cos}^2\theta - \frac{\partial\,C}{\partial\,x}\,\mathrm{sin}\theta\,\mathrm{cos}\theta\right)\,\vec{e_y} \quad[\mathrm{kg}\,\mathrm{m}^{-1}\,\mathrm{s}^{-1}],
\end{equation}

\noindent where $\theta$ is the wind angle relative to the $x$-axis of our grid. $C$ is again either generated via Eq. \ref{eq:plume_generation_equation} or calculated from TROPOMI satellite observations of methane. Eq. \ref{eq:F_dif_xy} is derived under the assumption that diffusion only takes perpendicular to the wind vector $\vec{w}$. If, however, we choose to ignore this assumption (but still assume that $K$ is constant in space), that we can work directly in the $(x,\,y)$ grid and state that 

\begin{equation} \label{eq:cartesian_F_dif}
    \vec{F}^\mathrm{dif} = K\,\frac{\partial\,C}{\partial\,x}\vec{e_x} +  K\,\frac{\partial\,C}{\partial\,y}\vec{e_y}
\end{equation}

\noindent Eqs. \ref{eq:simple_emission_equation}, \ref{eq:F_adv_xy}, \ref{eq:corrected_emission_equation}, \ref{eq:F_dif_xy}, and \ref{eq:cartesian_F_dif} are derived in S2. \\

We need to calculate spatial derivatives over a cartesian grid to fully obtain $E$ in Eqs. \ref{eq:simple_emission_equation} and \ref{eq:corrected_emission_equation}. For first derivatives, we use the fourth-order central finite difference \cite{Bronstein_1981}

\begin{equation}\label{eq:first_derivative}
    \frac{\partial\,V}{\partial\,p}\rvert_{p=i} = \frac{V\rvert_{p=i-2} - 8\,V\rvert_{p=i-1} + 8\,V\rvert_{p=i+1} - V\rvert_{p=i+2}}{12\,d}
\end{equation}

\noindent where $V$ is a spatially varying quantity and $d$ is the grid spacing in coordinate $p$. This numerical recipe is commonly used for emission estimates via the divergence method \cite{Beirle_2019, Veefkind_2023}. For second derivatives, we use the fourth order discretization \cite{Gibou_2005} 

\begin{equation}\label{eq:second_derivative}
    \frac{\partial^2\,V}{\partial\,p^2}\rvert_{p=i} = \frac{-\frac{1}{12}V\rvert_{p=i-2} + \frac{4}{3}\,V\rvert_{p=i-1} - \frac{5}{2}\,V\rvert_{p=i} + \frac{4}{3}\,V\rvert_{p=i+1} - \frac{1}{12}V\rvert_{p=i+2}}{d^2}.
\end{equation}

\subsection{Calculating time-averaged emission rates}\label{sec:e1_e2}
We calculate time-averaged estimated emission fields using two methodologies. In the first (denoted by $\overline{E}_1$), we calculate daily estimated emission fields and time average them to obtain $\overline{E}_1$. In the second methodology (denoted by $\overline{E}_2$), we time-average daily fluxes of $C$, and take the divergence of the time-averaged flux to obtain $\overline{E}_2$. With spatially complete observations of $C$ over an entire time period, the two methods yield identical results, but for TROPOMI observations of methane, data is often spatially masked due to cloud cover and albedo effects. Detailed equations describing these two methodologies is given is S4.

\subsection{Estimating the constant of turbulent diffusion K}\label{sec:estimating_K}
It is in practice difficult to estimate constants of turbulent diffusion. If $K$ is assumed to be constant in space and time, then the standard deviation ``width" of a Gaussian plume can be described via

\begin{equation}
    \sigma^2 = \frac{2\,K\,x}{u} \quad [\mathrm{m}^2],
\end{equation}

\noindent where $\sigma$ is the width of the plume $[\mathrm{m}]$, $x$ is the downwind distance in the plume $[\mathrm{m}]$, $u$ is the wind speed [$\mathrm{m}\,\mathrm{s}^{-1}$] and $K$ is the constant of turbulent diffusion [$\mathrm{m}^2\,\mathrm{s}^{-1}$] \cite{Stockie_2011, Seinfield_1998}. We take multiple GHGSat scenes of isolated methane plumes \cite{Varon_2019} and measure values of $\sigma$ at multiple downwind locations within each plume. We then fit a linear function to $\sigma^2$ against $x$ for each plume using the method of least squares. The slope of the fitted function yields $2\,K/u$, and thus $K$ can be determined as $u$ is known for each scene. We determine using these plumes that $K$ can vary between 10 and 400 $\mathrm{m}^2\,\mathrm{s}^{-1}$.


\section*{Acknowledgements}

\blackout{C R acknowledges financial support from Shell Research Ltd through the Cambridge Centre for Doctoral Training in Data Intensive Science grant number ST/P006787/1. For the purpose of open access, C R has applied a Creative Commons Attribution (CC BY) licence to any Author Accepted Manuscript version arising from this submission. C R thanks Dr. J P Veefkind at TU Delft for insightful discussion and advice. C R thanks the referees for their time and expertise in reviewing the manuscript.}

\section*{Author contributions}

\blackout{C R retrieved all data, wrote all code and conducted the analysis of this work. O S and D R supervised the project. R I guided the analysis. M J, P J, K M, and B H contributed discussions of the data, satellites and meteorological context.  All authors reviewed the manuscript.}

\section*{Conflict of interest}

We report no competing interests.

\section*{}

\bibliography{bibliography}

\newpage

\end{document}